# Memristive Behavior and Mechanism in Solid-State Nanopores

**Zhiwei Li[1], Ngan Hoang Pham[1,2]*, Shi-Li Zhang[1], Chenyu Wen[1]***

[1] *Division of Solid-State Electronics, Department of Electrical Engineering, Uppsala University, SE-751 21 Uppsala, Sweden*

[2] *Myfab Uppsala, Uppsala University, SE-751 21 Uppsala, Sweden*

*****Corresponding author:

ngan.pham@angstrom.uu.se

chenyu.wen@angstrom.uu.se

**Abstract**

Nanofluidic memristors whose conductance evolves through history-dependent ionic transport and dynamic interfacial processes are promising building blocks for ionic neuromorphic applications. However, most existing designs rely on biological nanopores, polymers, and two-dimensional materials, which limit scalable fabrication and poses challenges to integration of ionic computing circuits and systems. Here, we report memristive behaviors of silicon-based solid-state nanopores (SSNPs) fabricated based on wafer-scale semiconductor processes. The SSNPs exhibit hysteretic current-voltage characteristics with a dependence on voltage sweeping frequency, electrolyte concentration, and nanopore geometry. To investigate the physical origin of their memory feature, the measured current of the SSNPs is decomposed into resistive, capacitive, and memristive components. An ion adsorption-desorption kinetics is developed to explain and predict the memristive behavior. A dynamical system analysis further reveals that the memristive behavior arises from delayed relaxation, thereby linking the measured hysteresis to the observed adaptive ionic response. Together, these findings establish native SSNPs as scalable ionic memristive elements and provide a generalized electrokinetic mechanism for memristive behavior under nanoconfinement. The resulting analytical framework connects device characterization with the underlying dynamics, deepens mechanism understanding, and guides the design of ionic neuromorphic devices.

## 1. Introduction

Neuromorphic computing aims to replicate the signal processing tactics of biological nervous systems using physical devices[1–3]. In electronic hardware, memristors have attracted broad attention because their resistance can be programmed by exploiting the dependence of historical electrical stimulations, thereby allowing them to emulate synaptic plasticity, short-term memory, and state-dependent signal processing[4,5]. In biological systems, however, information processing is not purely based on electronic dynamics[6]. It is instead critically dependent on ion transport and dynamic interfacial processes. Its ionic origin has motivated the development of ionic neuromorphic devices, in which ions rather than electrons serve as the primary information carrier[7]. Nanopore/nanochannel devices have been explored for ionic memory realized by different material systems that exhibit history-dependent conductance caused by various mechanisms, such as ion clustering, ion-pairing, and transient ion concentration polarization[8–10]. Representative examples include chemical–electrical transduction in polyelectrolyte-confined pores[11], long-term programmable memory in Ångström-scale two-dimensional channels[12,13], and concentration-polarization-based memory in conical and ion-selective channels[8]. One bottleneck with these approaches, though, is wafer-scale fabrication and circuit-system integration[1,14–16].

Another bottleneck concerns mechanism clarity. In nanopores, the measured current is usually not a pure memristive signal. It can contain at least three coupled contributions: an ohmic or quasi-ohmic resistive current through the pore, a capacitive current from the membrane and chip structure, and a dynamic surface-related current[17]. Without clearly identifying these components, the interpretation of hysteretic current-voltage (*I*–*V*) curves can be ambiguous. Especially for the capacitive component, it can be easily confused with memcapative behavior. Unfortunately, there is lack of standardized analysis method for hysteresis characterization of nanofluidic memristors. More fundamentally, mechanisms pertaining to the memristive behavior are not well studied and correlation of the behaviors to their physical origins remain poor. Only a few studies have attempted to address this issue. In one such investigation, electrolyte composition, *p*H level, channel material, confinement, and driving frequency are identified to produce different hysteresis types in the device[12,13]. These findings indicate that ionic memristive behavior is inherently system- and condition-dependent. This ambiguity is the central problem we will address in this work.

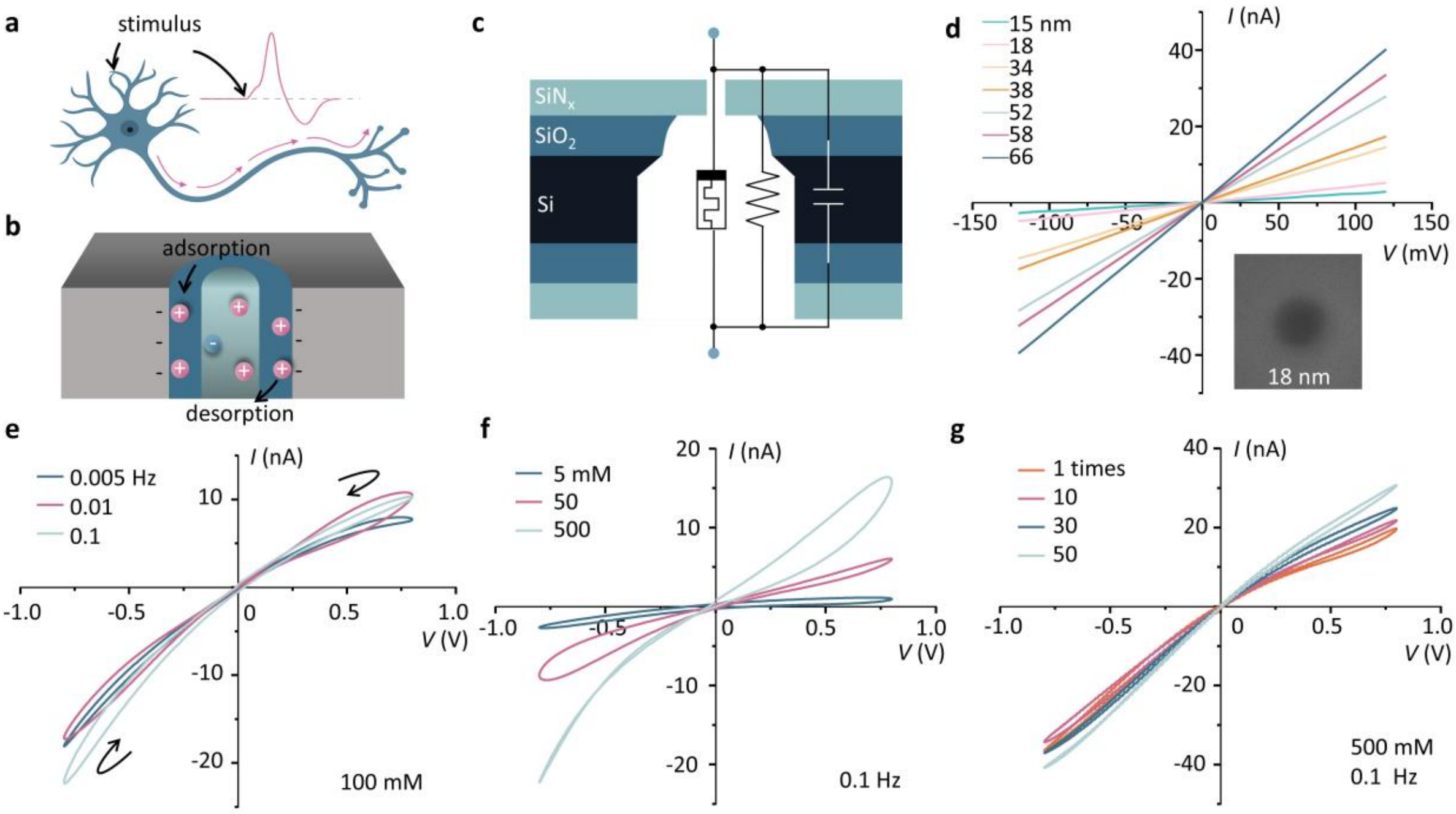

**Fig. 1**. Memristive ion transport in SSNP. **a**, Schematic illustration of neuronal spike generation and signal propagation. **b**, Conceptual representation of ion adsorption and desorption at the charged nanopore surface. **c**, Cross-sectional schematic of the SSNP and its equivalent electrical representation. **d**, I–V characteristics of nanopores with different diameters, with an SEM image of a representative 18 nm pore shown in the inset. **e**, Frequency-dependent hysteretic I–V responses measured in 100 mM KCl. **f**, Concentration-dependent hysteretic I–V responses measured at 0.1 Hz. **g**, I–V responses recorded during 50 consecutive voltage cycles in 500 mM KCl at 0.1 Hz.

Here, memristive behavior of solid-state nanopores (SSNPs) in a $SiN_x/SiO_2/Si$ structure (Fig. 1) is studied. The devices show hysteretic *I*–*V* curves over a range of electrolyte concentrations and voltage sweeping frequencies, while maintaining stable responses during repeated cycles. A physical model that decomposes the measured current into distinct contributions is developed. First, the resistive component is described by the sum of bulk and surface conductance. Second, the capacitive component is isolated as the parallel free-standing membrane and substrate-supported dielectric pathways. Third, the remaining hysteresis is attributed to dynamic ion adsorption-desorption on the SSNP wall, modeled using isothermal adsorption-desorption kinetics[18]. This decomposition allows for connecting the measured memristive responses to surface-charge dynamics and further predicting how hysteresis area changes with concentration and frequency. This physical model naturally purposes an analysis method of the hysteresis measurement of nanofluidic memristors, as a generalized analysis framework that can be used for different memristive devices. The memristive origin from a dynamical system perspective is further performed. Fixed-point and relaxation-time analyses show that the memory originates from delayed evolution toward a unique voltage-dependent stable state. The same dynamics generate gradual potentiation- and depression-like responses under consecutive voltage pulses. Together, these results establish a semiconductor-compatible SSNP platform and provide a general electrokinetic framework for interpreting characterization data and guiding the design of ionic memristive devices.

## 2. Results and Discussion

### 2.1. Memristive behavior in SSNPs

Our SSNPs were fabricated in a thin $SiN_x$ membrane supported by a $SiO_2/Si$ substrate using a wafer-scale semiconductor process, as detailed in the Methods section. More than 200 nanopores were fabricated on a single 4-inch silicon wafer with a yield exceeding 70%, as detailed in the Supporting Information (SI), Fig. S4. The cross-sectional schematic in Fig. 1c illustrates the device structure of an SSNP in which ionic transport occurs through the nanoscale pore. The fabrication and size control of the nanopore were first verified by electrical and structural characterization. The scanning electron microscope (SEM) image in the inset of Fig. 1d confirms the nanoscale pore opening. The nanopore was immersed in a KCl electrolyte for electrical measurements. Under a bias voltage across the membrane, ions transport through the pore and form an ionic current. The *I*–*V* measurements performed show an approximate linear increase in ionic current *I* with applied voltage *V*. A higher conductance is expected for a larger pore. These results indicate that the measured current originates from the well-defined nanopores rather than from membrane leakage or uncontrolled defects (Fig. S3). The pore diameter and surface charge density are further evaluated using the conductance model[19]. The extracted pore sizes agree well with the designed dimensions, while the fitted surface charge densities fall within the range reported for silicon-based nanopore surfaces.

After confirming the nanopore fabrication, the dynamic ionic response under periodic voltage excitations was examined. When a sinusoidal voltage was applied across the nanopore, the current exhibited a hysteretic *I*–*V* trajectory rather than an ohmic response. As shown in Fig. 1e, the loop shape depends strongly on the voltage sweeping frequency. At lower frequencies, the hysteresis becomes more pronounced, whereas increasing the frequency reduces the loop opening. This frequency dependence indicates that the conductance is governed by an internal ionic state that evolves on a finite timescale and cannot respond instantaneously to the applied voltage. The *I*–*V* curves exhibit non-zero-crossing hystereses with a narrow waist near zero voltage, rather than a strictly pinched loop at the origin. This current at zero applied voltage is an important indication of the current containing not only the memristive component but also resistive and capacitive contributions. Therefore, the apparent *I*–*V* loop in Fig. 1e should be regarded as the total electrical response of the device. The intrinsic memristive contribution is extracted and analyzed in later sections after isolating the resistive and capacitive components.

The voltage response also shows a clear dependence of the current loop on electrolyte concentration. At the same sweeping frequency, increasing KCl concentration substantially enlarges the current amplitude and the hysteresis loop, as shown in Fig. 1f. This behavior is consistent with the ionic transport in a surface-charged nanopore; both the conductivity of the electrolyte and the surface conductance vary with electrolyte concentration. The concentration-dependent loop evolution suggests that the memory effect is closely related to ion–surface interactions rather than being caused by parasitic capacitance or instrumental delay.

The stability of the response was subsequently evaluated by applying 50 consecutive voltage sweeping cycles under identical conditions (Fig. 1g). The *I*–*V* curves largely overlap throughout the repetitions, demonstrating that the nanopore retains a reproducible memristive response without obvious degradation. Together, these observations establish the essential experimental observation that native SSNPs based on Si-related membrane materials and compatible with semiconductor fabrication processes are characterized by reproducible ionic memristive behavior avoiding biological channels, polymer deformation, two-dimensional material assembly, or additional chemical functionalization.

## 2.2. Decomposition of the SSNP electrical response

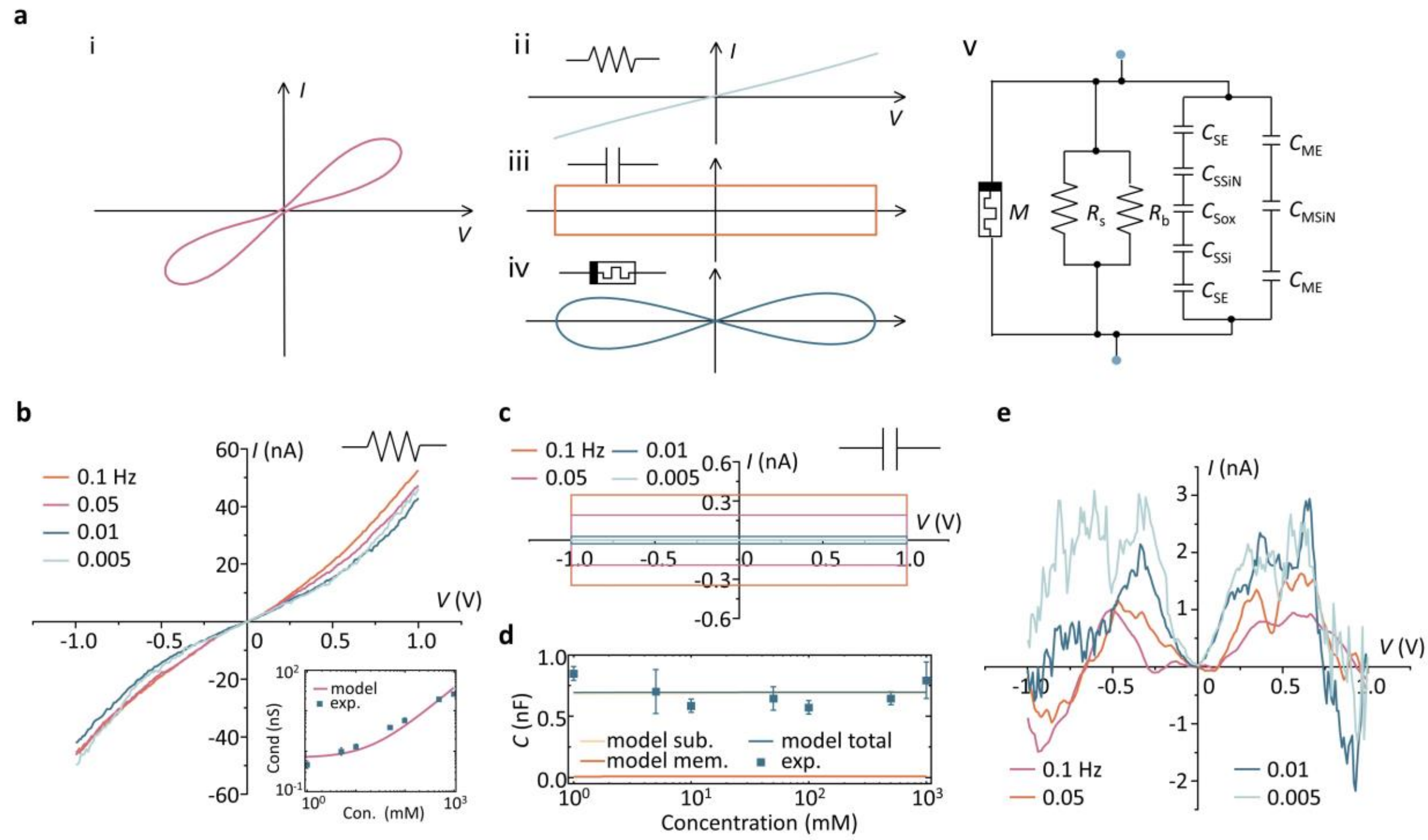

**Fig. 2**. Decomposition of the SSNP electrical response. **a**. Conceptual decomposition of the experimental hysteretic *I–V* response in (**i**) into resistive (**ii**), capacitive (**iii**), and adsorption-based memristive (**iv**) contributions, together with the corresponding equivalent circuit in (**v**). **b**, Extracted resistive response under triangular voltage excitation at different frequencies. The inset compares experimental conductance with the combined bulk and surface conduction model. **c**, Capacitive *I–V* response at different excitation frequencies. **d**, Experimentally extracted capacitance as a function of KCl concentration and the modeled capacitor. **e**, Adsorption-based hysteresis after subtraction of the resistive and capacitive contributions.

The hysteretic response measured on an SSNP includes contributions from steady ionic conduction, dielectric charging, and interfacial dynamics (see Fig. 2a for a schematic summary). These contributions must be separated before the physical origin of the memory effect can be identified. The original current hysteretic response, $I_{\mathrm{meas}}(t)$, in Fig. 2a(i) is represented as the superposition of the resistive response, $I_{\mathrm{R}}$, in Fig. 2a(ii), the capacitive response, $I_{\mathrm{C}}$ , in Fig. 2a(iii), and the memristive response, $I_{\mathrm{M}}$, in Fig. 2a(iv). Thus,

$$I_{\mathrm{meas}}(t) = I_{\mathrm{R}}(t) + I_{\mathrm{C}}(t) + I_{\mathrm{M}}(t) \tag{1}$$

The physical basis of this decomposition is summarized by the equivalent circuit in Fig. 2a(v). The decomposition was facilitated using triangular voltage sweeps with which the voltage changes at a constant rate along each cycle. Firstly, the currents from the increasing- and decreasing-voltage branches were averaged at each voltage point. The resulting mean current is the resistive contribution:

$$I_R(V) = \frac{I_{\uparrow}(V) + I_{\downarrow}(V)}{2} \tag{2}$$

Secondly, subtracting this component from the original response yields a residual response containing the capacitive and memristive currents. Because the capacitive current should remain constant during each linear voltage ramp, it is determined from the residual current at $V = 0$.

$$I_{\mathrm{C}}(t) = I_{\mathrm{meas}}(V = 0) \tag{3}$$

Thirdly, the remaining memristive current is then obtained from

$$I_{\mathrm{M}}(t) = I_{\mathrm{meas}}(t) - \frac{I_{\uparrow}(V) + I_{\downarrow}(V)}{2} - I_{V=0} \tag{4}$$

The complete branch-resolved curves and intermediate processing steps are provided in Supplementary Note 1 in the SI. This framework splits the measured response according to branch symmetry and sweep-rate dependence before assigning microscopic mechanisms. It therefore provides a general approach for analyzing hysteretic signals in ionic and electrochemical devices whose currents comprise resistive, capacitive, and memristive contributions. The physical origins of each extracted component for the SSNPs is examined individually using the corresponding physical models.

The extracted resistance is first examined. As shown in Fig. 2b, the extracted component is approximately linear near zero voltage but develops a noticeable curvature at higher voltages.

Such deviations from ideal ohmic conduction can be attributed to field-induced charge redistribution and ion concentration polarization near the pore entrances[20]. The resistive component can be modeled as the sum of bulk and surface conductance as follows:

$$G = G_\mathrm{b} + G_\mathrm{s} = \frac{\pi {D_p}^2}{4L_\mathrm{eff}} q N_\mathrm{A} (\mu_{\mathrm{K}^+} + \mu_{\mathrm{Cl}^-}) c + \frac{\pi D_p}{L_\mathrm{eff}} \mu_{\mathrm{K}^+} |\sigma_0| \quad (5)$$

in which $D_p$ is the nanopore diameter, $L_\mathrm{eff}$ the effective pore length[21], $q$ elementary charge, $N_\mathrm{A}$ Avogadro constant, $\mu_{\mathrm{K}^+}$ and $\mu_{\mathrm{Cl}^-}$ respectively the electrophoretic mobility of potassium and chloride ions, $c$ the electrolyte concentration, and $\sigma_0$ the intrinsic surface charge density. The inset of Fig. 2b shows that the model, Eq. 5, reproduces the concentration-dependent conductance experimentally extracted. Bulk transport dominates at high concentrations, whereas counterion enrichment and surface conduction become increasingly important at lower concentrations. These trends are consistent with established electrokinetic descriptions of charged nanopores[21]. From the fitting, the diameter of nanopore, $D_p$, and surface charge density, $\sigma_0$, are extracted as $13\ nm$ and $-0.012\ \mathrm{C/m^2}$ for the device shown in Fig 2.

After removal of the resistive background, the finite zero-voltage residual was used to reconstruct the capacitive response shown in Fig. 2c. The approximately rectangular *I–V* profile follows directly from the constant sweep rate of the triangular waveform. The current reversal near sweeping turning points is experimentally more complex because the sweep changes direction over a finite time and may be accompanied by a rapid interfacial relaxation. These localized turning-point transients were therefore ignored in the capacitive current. The capacitive assignment is further supported by the linear dependence of current on sweeping frequency, as shown in Fig. S5. The physical origin of the extracted capacitance is associated with two regions in our devices, *i.e.*, the substrate-supported area, $C_\mathrm{s}$, and the free-standing membrane, $C_\mathrm{M}$, shown in Fig. 2a(v). The pathway of substrate area comprises the electrical double layers (EDL), the $SiN_x$ and $SiO_2$ layers, and the silicon capacitance, whereas the membrane region pathway contains the $SiN_x$ membrane and the two EDLs on its upper and lower surface. These contributions act in parallel, yielding,

$$C_\mathrm{total} = C_\mathrm{S} + C_\mathrm{M} \quad (6)$$

The complete dielectric-stack model is presented in Fig. S5 in the SI. The calculated capacitances agree well with the experimentally extracted values over the investigated concentration range, as shown in Fig. 2d. The substrate-supported region provides the dominant contribution to the total capacitance, whereas the contribution from the considerably smaller free-standing membrane is comparatively minor.

Subtracting both resistive and capacitive components yields the memristive residual shown in Fig. 2e. This component increases with decreasing the excitation frequency and cannot be explained by quasi-static conduction or chip capacitance. The decomposition itself does not determine its microscopic origin either. Instead, it presents the voltage response that must be accounted for by a state-dependent physical process.

### 2.3. Adsorption model for memristive behavior in SSNPs

The frequency dependence of the memristive current (Fig. 2e) indicates that the internal ionic state requires a finite time to evolve under the applied electric field. Large fluctuations are observed near both positive/negative voltage peaks around which transient capacitive currents arise during reversal of the voltage sweep. The subsequent quantitative analysis is therefore restricted to the range from −0.8 to 0.8 V, which retains the principal hysteretic features while excluding the turning-point regions most strongly affected by capacitive transients.

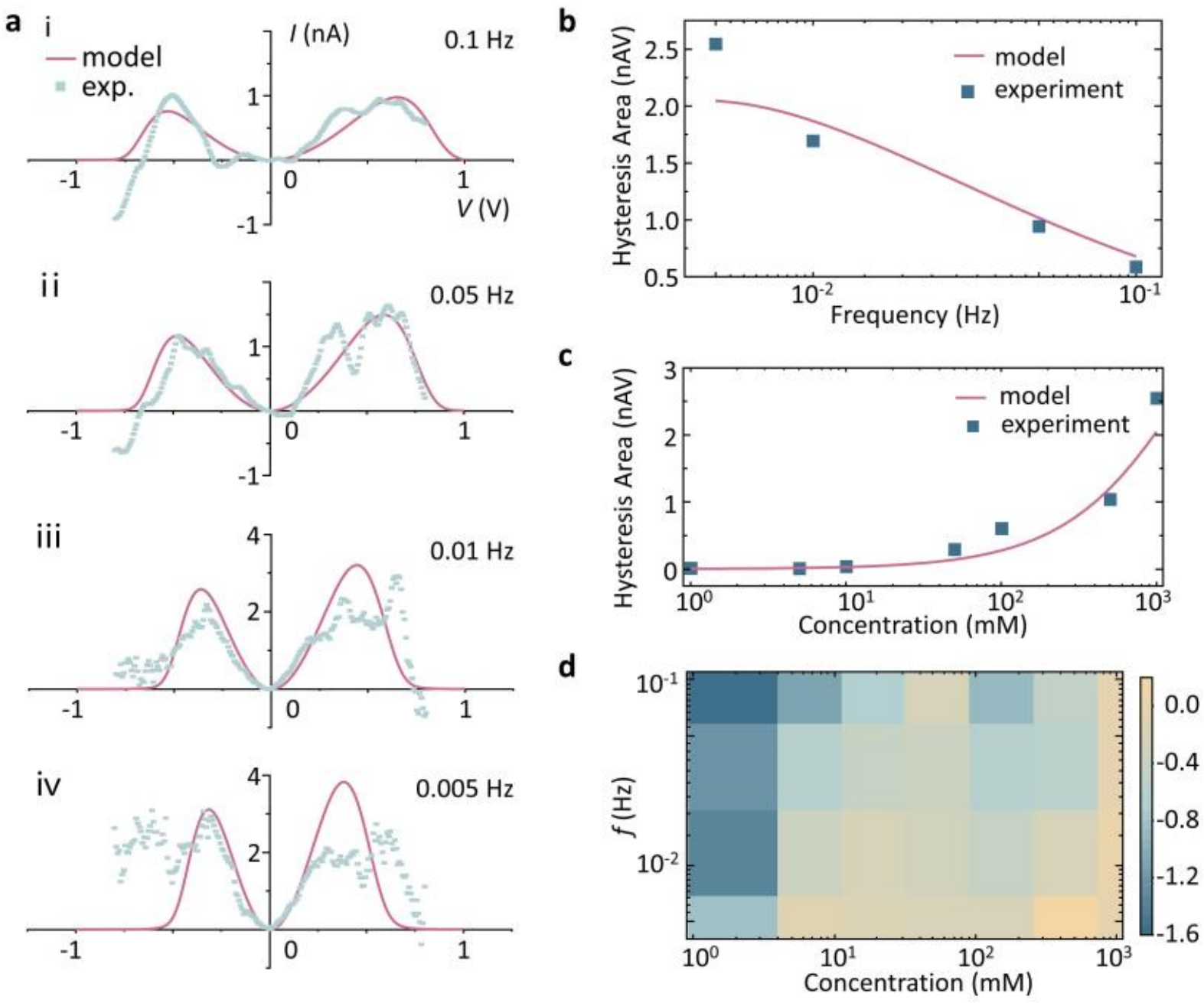

**Fig. 3**. Validation of the adsorption-based hysteresis model. **a**, Comparison between experimental and modeled hysteresis currents at (**i**) 0.1 Hz, (**ii**) 0.05 Hz, (**iii**) 0.01 Hz, and (**iv**) 0.005 Hz. **b**, Experimental and modeled hysteresis areas as a function of excitation frequency. **c**, Experimental and modeled hysteresis areas as a function of KCl concentration. **d**, Map of the discrepancy between modeled and experimental hysteresis areas over the investigated concentration–frequency space.

To establish the model, it is proposed that hysteresis is caused by an electric field-dependent interfacial adsorption state. Hydrated $SiN_x$ surfaces have been described using reactive silanol- and amine- groups whose charge states depend on the chemical environment[22]. Experiments on solid-state nanopores have directly connected current fluctuations to the kinetics of surface protonation and deprotonation[23]. Theoretical and experimental studies further show that the surface charge density of nanopores varies with electrolyte concentration and fabrication process[22]. In parallel, molecular dynamics simulations of ion-binding nanopores demonstrate that a delayed occupation of interfacial sites can couple the internal ionic state to pore conductance and produce a memristive response[24]. Complementary kinetic and atomistic studies describe this behavior in terms of transient ion binding and thermally activated barrier crossing, providing the physics basis for the time-dependent surface state adopted in our model[22].

Below, the adsorption/embedding of ions on the nanopore surface is considered. The coverage of the adsorption active sites can be expressed as:

$$\theta(t) = \frac{\Gamma(t)}{\Gamma_{max}} \tag{7}$$

where, $\Gamma(t)$ is the instantaneous ion adsorption density at time $t$ and $\Gamma_{max}$ the maximum density of effective adsorption sites. The corresponding surface charge density is thus

$$\sigma(t) = \sigma_0 + zqN_A\Gamma_{\mathrm{max}}\theta(t) \tag{8}$$

where, $\sigma_0$ is the intrinsic surface charge density extracted from the resistive model and $z$ the valence associated with the adsorbed ionic species. It is worth noting that the adsorption variable used here is an effective surface state, which may represent reversible counterion association, changes in the protonation state of reactive groups, or local reorganization of the EDL. Our model does not resolve these microscopic processes individually but considers an ensemble of the overall process effectively.

The evolution of the surface coverage can be described as

$$\frac{d\theta}{dt} = k_{\mathrm{ad}}(E)c[1-\theta] - k_{\mathrm{des}}(E)\theta \tag{9}$$

where, $k_{\mathrm{ad}}(E)c$ and $k_{\mathrm{des}}(E)$ are, respectively, the effective adsorption and desorption rates. Physically, ions reach and leave the interfacial region through Brownian motion driven by its thermal energy. Both adsorption and desorption require the ions crossing their respective energy barrier. The rate constants, $k_{\mathrm{ad}}c$ and $k_{\mathrm{des}}$ , are proportional to the respective probability of crossing the energy barrier. An applied electric field, $E = V/L_{eff}$, alters this energy landscape and changes the barrier height. These considerations lead to an exponential electric field dependence of the rate constants as follows:

$$k_{\mathrm{ad}}(E) = k_{\mathrm{ad,0}}e^{\beta_{\mathrm{a}}|E|+\beta_{\mathrm{s}}E} \tag{10}$$

$$k_{\mathrm{des}}(E) = k_{\mathrm{des,0}}e^{-\gamma_{\mathrm{a}}|E|-\gamma_{\mathrm{s}}E} \tag{11}$$

The terms containing $|E|$ is the common-mode component, describing changes in the effective barrier that depend on field amplitude and are therefore common to both voltage polarities. They can phenomenologically include field-enhanced interfacial polarization, ionic focusing, and changes in the probability of reaching an adsorption configuration. The terms containing signed $E$ is the differential-mode component, which accounts for polarity-dependent effects arising from unequal entrance and exit conditions, nonuniform surface chemistry, asymmetric interfacial polarization, or differences between the two membrane surfaces. The local field near a nanopore is spatially nonuniform, particularly in the access regions. Hence, $E = V/L_{\mathrm{eff}}$ should be interpreted as an effective axial electric field. Accordingly, $\beta_{\mathrm{a}}$, $\beta_{\mathrm{s}}$, $\gamma_{\mathrm{a}}$, and $\gamma_{\mathrm{s}}$ are effective field-sensitivity coefficients (see Supporting Note 2 for more details). Studies of short nanopores have similarly shown that the charged exterior membrane surfaces and pore entrances can influence the measured ionic transport[25].

Changes in surface coverage modify the ionic transport through the surface-conductance contribution,

$$G_{\mathrm{s}}(\theta) = \frac{\pi D_p}{L_{\mathrm{eff}}}\mu|\sigma(t)| \tag{12}$$

and, thus, causes the memristive component of the ionic current:

$$I_{\mathrm{M}}(t) = \left[G_{\mathrm{s}}(\theta) - G_{\mathrm{s},\sigma_0}\right]V(t) \tag{13}$$

The effective charge represented by $\sigma(t)$ describes the net interfacial state that contributes to the measured ionic conductance, including the inner pore wall and, where relevant, the adjacent membrane surfaces within the access region. Because $\theta$ cannot instantaneously follow the changing electric field, the increasing- and decreasing-voltage branches can have different surface coverages at the same voltage. The measured hysteresis is therefore interpreted as the electrical consequence of delayed surface-charge regulations rather than as a direct measurement of the adsorption itself.

The kinetic parameters were calibrated by fitting the $I_M$-$t$ current curve, measured in 1M KCl. There are in total 7 independent parameters in the model, *i.e.*, $\Gamma_{\mathrm{max}}$, $k_{\mathrm{ad,0}}$, $k_{\mathrm{des,0}}$, $\beta_{\mathrm{a}}$, $\beta_{\mathrm{s}}$, $\gamma_{\mathrm{a}}$, and $\gamma_{\mathrm{s}}$. A same values of parameters are used across the same device, by taking into account the strong individual differences of nanopore devices since the ionic current is strongly dependent on the local physicochemical conditions inside and near a nanopore[26]. It is worth noting that some parameters are confined within a physically reasonable range during the fitting. For example, $\Gamma_{\mathrm{max}}$ is in the range of $10^{-6}$ to $10^{-5}$ mol/m$^2$ [27]. No frequency-specific parameter is introduced, but only the period of the applied waveform is changed. The resulting model with the experimental hysteresis currents at 0.1, 0.05, 0.01, and 0.005 Hz are compared in Fig. 3a. The model reproduces the principal voltage dependence, polarity asymmetry, and increase in residual current at lower frequencies. This frequency dependence arises because, at lower frequencies, the surface state has a sufficient time to evolve toward its field-dependent equilibrium state, producing a large difference in $\theta$ between the increasing- and decreasing-voltage branches. At higher frequencies, the available response time is shorter, limiting the change in $\theta$ during each voltage sweeping branch and thereby reducing the hysteresis current.

The frequency dependence is quantified using the integrated absolute residual over the selected voltage interval as an index,

$$A_{\mathrm{hys}} = \int_{-V_{\mathrm{int}}}^{V_{int}} |I_{M\uparrow}(V)|\, dV \tag{14}$$

where $V_{int}$ denotes the voltage range used for quantitative analysis and was set to 0.8 V in this study to exclude the transient effects from the voltage turning at ±1 V. As shown in Fig. 3b, both experiment and model exhibit a monotonic reduction in $A_{\mathrm{hys}}$ with increasing frequency over the investigated range. At low frequencies, adsorption and desorption have sufficient time to produce a substantial difference between the surface states sampled on the two voltage branches. At higher frequencies, $\theta$ changes only slightly during one cycle, thereby reducing both the conductance difference and the integrated hysteresis. This trend applies to the measured frequency window. In the quasi-static limit, a sufficiently slow driving field would ultimately allow the surface state to remain close to its instantaneous stable fixed point and could again reduce the hysteresis (see discussion in Section 2.4 for more details).

Varying the electrolyte concentration provides an independent validation of our model. For this comparison, the adsorption-site density, $\Gamma_{\mathrm{max}}$, intrinsic rate constants, $k_{\mathrm{ad,0}}$, $k_{\mathrm{des,0}}$, and field-sensitivity coefficients, $\beta_{\mathrm{a}}$, $\beta_{\mathrm{s}}$, $\gamma_{\mathrm{a}}$, and $\gamma_{\mathrm{s}}$, are all kept the same values as those in the reference condition, while the electrolyte concentration, *c*, is changed explicitly in the adsorption flux $k_{\mathrm{ad}}(E)c[1-\theta]$. The hysteresis area is shown in Fig. 3c to increase with KCl concentration. A higher concentration offers more supply of ions at the interfacial region, increases the probability of adsorption, and enlarges the dynamic modulation of surface charge during a voltage cycle. It is worth noting that this trend is observed after the subtraction of the conductive background, and therefore, cannot be explained by an increase in bulk ionic conductivity at higher concentrations.

The discrepancy of hysteresis area, $A_{\mathrm{hys}}$, between the model and experimental results across all the measured concentration–frequency combinations is summarized in Fig. 3d. The model captures the dominant effects of both excitation timescale and ionic availability using the same kinetic structure. The largest relative discrepancies occur where the extracted memory current is small or where the uncertainty from capacitive subtraction, baseline drift, and current noise becomes comparable to the residual signal. The agreement across the waveform shape, frequency, and concentration provides a strong support for a field-dependent adsorption state. The principal conclusion is therefore that a single adsorption-like surface variable is sufficient to account for the major features of the isolated hysteresis.

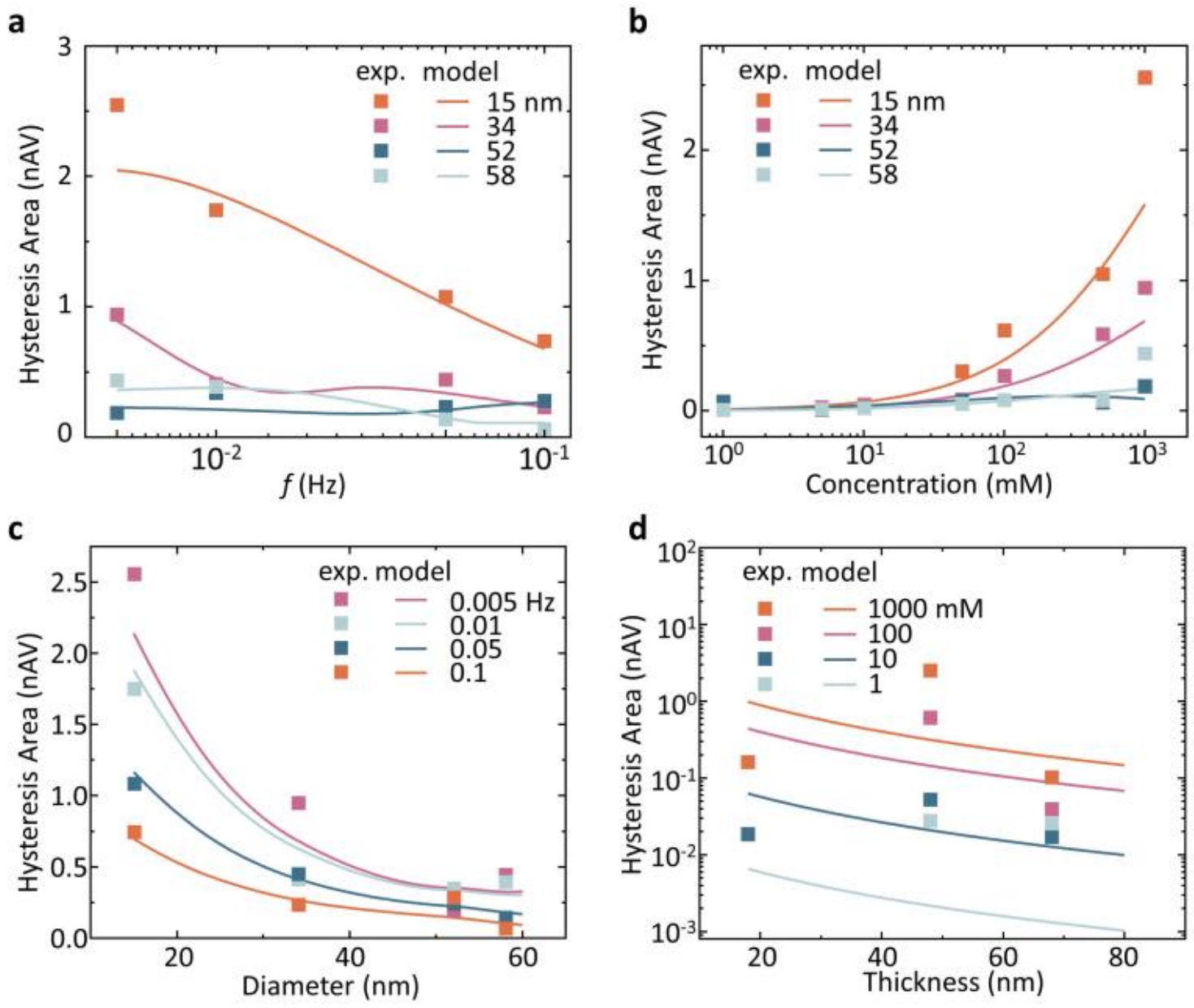

**Fig. 4**. Geometric dependence of adsorption-based hysteresis in SSNP. **a**, Experimental and modeled hysteresis areas as a function of excitation frequency for nanopores with different diameters. **b**, Experimental and modeled hysteresis areas as a function of KCl concentration for different pore diameters. **c,** Dependence of hysteresis area on pore diameter at different excitation frequencies. **d**, Dependence of hysteresis area on $SiN_x$ membrane thickness at different KCl concentrations.

The geometry variation in Fig. 4 provides a further assessment of this physical picture and demonstrates the power of the prediction ability of the model. For each device, the interfacial site density, $\Gamma_{\mathrm{max}}$, and kinetic coefficients, $k_{\mathrm{ad},0}$, $k_{\mathrm{des},0}$, $\beta_{\mathrm{a}}$, $\beta_{\mathrm{s}}$, $\gamma_{\mathrm{a}}$, and $\gamma_{\mathrm{s}}$, were held constant, while the measured pore geometry was incorporated in the model. The memristive component, *i.e.*, $A_{\mathrm{hys}}$, is generally stronger in smaller pores, it increases with concentration and decreases with excitation frequency, as shown in Fig. 4a-c, respectively. Reducing the pore diameter increases the surface-to-volume ratio in conductance and the relative significance of the surface transport component. The nanopore surface charge density has also been shown device-to-device differences in the confined regime, which may also depend on concentration and pore geometry[18]. These can be the major reasons for the scatter of experimental measurement results around the model prediction trends.

The model predicts a weaker hysteresis from a pore with thicker membrane as shown in Fig. 4d. At a fixed applied voltage, a larger $L_{\mathrm{eff}}$ reduces the effective axial field and therefore weakens the field-induced modulation of the adsorption and desorption rates. Increasing $L_{\mathrm{eff}}$ also reduces the surface-conductance contribution through its inverse geometric dependence. The experimental results exhibit an obvious dispersion and a large scatter, reflecting variations in pore profile, interfacial chemistry, and subtle effective adsorption-site density among the individual devices, which still poses a common challenge in SSNP devices[28]. Nevertheless, the principal dependences on frequency, concentration, pore diameter, and membrane thickness remain consistent with a memory response governed by electric field modulated surface adsorption kinetics.

### 2.4. Dynamical system interpretation

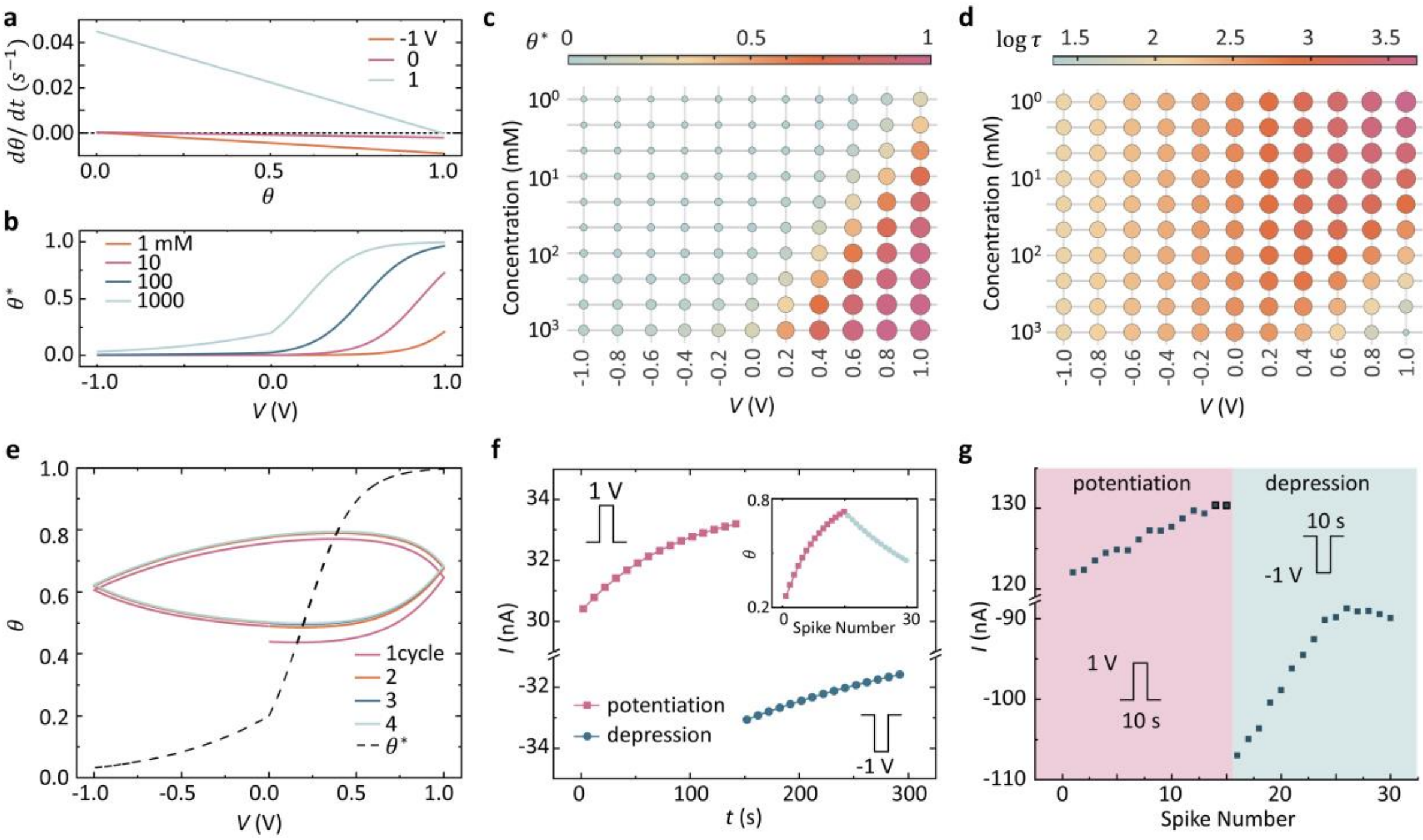

**Fig. 5**. Dynamical origin and pulse-driven adaptation of the SSNP response. **a**, Phase-line analysis of the surface-coverage dynamics, $d\theta/dt$ , at applied voltages of −1, 0, and 1 V. **b**, Voltage-dependent fixed point, $\theta^*$, at different KCl concentrations. **c**, Map of $\theta^*$as a function of voltage and electrolyte concentration. **d**, Corresponding relaxation-time map, expressed as $log\tau$. **e**, Simulated evolution of $\theta$ during successive triangular-voltage cycles, with the $\theta^*$ shown for reference. **f**, Simulated readout current under consecutive potentiation and depression pulses. The inset shows the corresponding evolution of $\theta$ with pulse number. **g**, Experimental readout current under repeated potentiation and depression pulses, demonstrating gradual and reversible ionic adaptation.

The adsorption model is a one-dimensional dynamical system in which the surface coverage $\theta(t)$ acts as an internal memory variable of the nanopore. This interpretation connects the experimentally observed *I*–*V* hysteresis to the time-dependent evolution of surface charge. This approach provides a physical basis for different memristive behaviors, including pulse-driven adaptation.

Regarding the adsorption dynamics model in Eq. 9, at each fixed voltage and concentration, the adsorption dynamics has a single fixed point given as,

$$\theta^* = \frac{k_{\mathrm{ad}}(V)c}{k_{\mathrm{ad}}(V)c + k_{\mathrm{des}}(V)} \tag{15}$$

Its local stability is determined by,

$$\frac{\partial}{\partial\theta}\left(\frac{d\theta}{dt}\right)\bigg|_{\theta^*} = -[k_{ad}(V)c + k_{des}(V)] < 0 \tag{16}$$

which indicates that the fixed point is stable under each fixed-voltage condition. Therefore, the present adsorption model does not rely on multi-stability. Instead, the memory effect originates from the finite relaxation of a stable surface state under time-dependent electrical stimulation.

This behavior is represented by the phase-line analysis in Fig. 5a. At representative voltages, $d\theta/dt$ varies linearly with $\theta$ and each curve crosses zero at a voltage-dependent fixed point. For $\theta < \theta^*$, adsorption dominates and increases the surface coverage, whereas for $\theta > \theta^*$, desorption drives the system back toward the fixed point. The voltage dependence of this stable state is shown in Fig. 5b. As the applied voltage is swept from negative to positive values, the fixed point $\theta^*$ increases continuously. For the polarity-dependent rate constants used here, positive bias shifts the surface toward an adsorption-dominated state, whereas negative bias favors a lower-coverage state. The fixed point also depends strongly on electrolyte concentration. Higher concentrations increase the adsorption driving term $k_{\mathrm{ad}}c$, thereby shifting the stable coverage toward larger values. This concentration dependence is important because it determines the possible range of surface charge density change during the electrical stimulation.

The combined influence of voltage and concentration is visualized in Fig. 5c as a map of fixed points $\theta^*(V, c)$. It shows, in fact, the adsorption equilibrium landscape of the device. At low concentrations, the surface coverage remains relatively small because the number of ions available for adsorption is limited. At higher concentrations, the stable coverage increases over a broader voltage range, indicating a larger possible modulation of surface charge density and surface conductance. This trend provides a direct explanation for the experimentally observed increase in hysteresis area with KCl concentration: higher concentrations do not merely increase the total ionic current; they also enlarge the dynamic range over which the surface state can evolve during voltage sweeping.

The relaxation time constant associated with this fixed point is given by,

$$\tau(V, c) = \frac{1}{k_{\mathrm{ad}}(V)c + k_{\mathrm{des}}(V)} \tag{17}$$

It determines how rapidly the surface state approaches $\theta^*$ from a new initial state, for example, after a change in voltage. The dynamics for a voltage step can be expressed as,

$$\theta(t) = \theta^* + [\theta(0) - \theta^*]e^{\left(-\frac{t}{\tau}\right)} \quad (18)$$

However, the adsorption state does not change instantaneously. Instead, it relaxes toward the voltage-dependent fixed point over a characteristic timescale of $\tau$. This finite response time is the origin of the hysteresis. During a triangular voltage sweep, $\theta(t)$ continuously attempts to follow $\theta^*(V, c)$, but because the voltage changes with time, the surface state lags behind the change of fixed point. As a result, the nanopore can have different surface coverage values at the same voltage during the forward and reverse sweeps, producing different surface conductance states and hence a hysteretic *I*–*V* loop.

The map in Fig. 5d shows the dependence of the relaxation time constant on voltage and concentration. The longest $\tau$ is found when the combined adsorption and desorption rates are low, especially under weak voltage stimulation and/or low adsorption availability. Increasing concentration generally reduces $\tau$ because the adsorption term $k_{\mathrm{ad}}c$ increases. However, a shorter $\tau$ does not by itself guarantee a larger $A_{\mathrm{hys}}$. The hysteresis is maximized when two conditions are satisfied simultaneously: the state of surface charge density changes by a sufficiently large amount and this change lags behind the applied voltage change. In our experimental range, increasing concentration enlarges the accessible change in $\theta$ and allows the surface state to evolve more appreciably, while still preserving a finite delay. This combination increases $A_{\mathrm{hys}}$. In the extreme of fast relaxation where $\tau$ is much shorter than the voltage changing time, the surface state would approach quasi-equilibrium and the $A_{\mathrm{hys}}$ will decrease. This dynamic also explains why hysteresis depends on voltage sweeping frequency. At high frequencies, the voltage changes too rapidly for the adsorption state to evolve accordingly. Hence, the variation in $\theta(t)$ during one voltage sweeping cycle is reduced. At lower frequencies, the state has more time to respond, resulting in a larger surface charge density modulation and stronger hysteresis. Therefore, the observed frequency dependence can be understood as a competition between the external voltage changing speed and the intrinsic adsorption relaxation speed (see Fig. 5e).

The present adsorption model can further predict adaptive behaviors under rectangular voltage pulses, as shown in Fig. 5f. During a positive writing pulse, the surface state is driven toward the fixed point under such positive-voltage,

$$\theta_+^* = \theta^*(V_{\mathrm{w}}, c) \quad (19)$$

with relaxation time,

$$\tau_+ = \tau(V_{\mathrm{w}}, c) \quad (20)$$

After a pulse of duration $T_{\mathrm{p}}$, the state becomes,

$$\theta_{\mathrm{after}} = \theta_+^* + (\theta_{\mathrm{before}} - \theta_+^*)e^{\left(-\frac{T_{\mathrm{p}}}{\tau_+}\right)} \quad (21)$$

During the following rest interval at zero voltage, the state partially relaxes toward the fixed point of zero bias,

$$\theta_0^* = \theta^*(0, c) \quad (22)$$

according to

$$\theta_{\mathrm{next}} = \theta_0^* + (\theta_{\mathrm{after}} - \theta_0^*) e^{\left(-\frac{T_{\mathrm{rest}}}{\tau_0}\right)} \tag{23}$$

Repeated positive pulses therefore gradually increase $\theta$, but the increment becomes smaller as the surface approaches saturation, which leads to a potentiation response. Conversely, negative erasing pulses drive the system toward the fixed point with lower-coverage, producing a depression behavior. The bounded adsorption/desorption coverage, $0 \leq \theta \leq 1$, gives rise to saturation, while relaxation during the interpulse intervals gives rise to volatility. Thus, the same adsorption/desorption dynamics that generate low-frequency *I*–*V* hysteresis also provide a physical mechanism for short-term adaptive ionic memory. In an experiment, application of 15 consecutive writing spikes followed by 15 erasing spikes produced gradual and reversible modulations of the read current, as shown in Fig. 5g, in agreement with the model-predicted adaptive adsorption dynamics (Fig. 5f). With the assistance of simulation, it is demonstrated that the measured update characteristics can be used as a synaptic behavior for learning and classification on the MNIST dataset, as an assessment of neuromorphic applicability (see Supporting Note 3 in the SI for more details).

## 3. Conclusions

Memristive behaviors in $SiN_X$ SSNPs have been demonstrated. The nanopores exhibit hysteretic *I*–*V* curves whose amplitude and area depend on electrolyte concentration and sweeping frequency of excitation voltage. The response remains stable over repeated measurements, indicating that the observed hysteresis is an intrinsic and reproducible feature of the nanopore. The SSNPs offer a large-scale fabricable solution of ionic neuromorphic devices, with the compatibility with standard semiconductor processing. All this may enable the CMOS device/circuit integration with the ionic neuromorphic system. In addition, a central contribution of this work is the physical decomposition of the measured current into resistive, capacitive, and memristive components. Hysteresis is captured by an adsorption-desorption model in which the dynamic surface coverage of adsorbed ions modulates the nanopore surface charge density and, thus, surface conductance.

The dynamical system analysis is employed to show that the surface coverage approaches a unique fixed point under given bias voltages. The observed memristive behavior therefore arises from finite relaxation of the adsorption state under time-dependent stimulations. This interpretation explains the potentiation and depression observed under repeated writing and erasing pulses. This interfacial kinetics, therefore, accounts for both the low-frequency *I-V* hysteresis and the pulse-driven adaptive response.

These findings establish a direct connection between the ion adsorption kinetics, nanopore surface conductance, and macroscopic memristive behavior. They also provide experimentally grounded design parameters for control of ionic memristive behavior through electrolyte concentration, voltage changing timescale, and nanopore geometry. Further improvements in membrane uniformity, surface physicochemical condition control, and device-to-device reproducibility will be required for circuit level implementation. In summary, the combination of scalable semiconductor fabrication and a physically interpretable memristive mechanism positions SSNPs as promising building blocks for future ionic synapses, adaptive sensors, and hybrid ion–electron neuromorphic systems.

## Methods

**Nanopore fabrication**. The SSNP fabrication process started with a double-side polished *p*-type prime silicon wafer with a thickness of 300 μm and a resistivity of $1 - 10\ \Omega \cdot \mathrm{cm}$. The wafer was first cleaned by RCA-1 and RCA-2 treatments at 70 °C for 10 min each, followed by a 1 min HF 2% dip to strip native oxide. A ~100-nm-thick $SiO_2$ pad oxide was then grown by dry thermal oxidation, followed by the growth of a ~50-nm-thick low-stress $SiN_x$ layer by means of low-pressure chemical vapor deposition. Nanoscale pore windows were defined on the front side of the wafer using electron-beam lithography (JEOL 8100 FS) with AR-P 6200.09 resist, followed by development. The patterns were then transferred to the $SiN_x$ layer by reactive-ion etching (RIE) (Advanced Vacuum Vision 320 Reactive Ion Etcher). After front-side SEM inspection (Merlin, Zeiss), the patterned surface was protected with S1813 photoresist. Large backside membrane windows were subsequently defined by AZ 10XT photolithography and opened up through the backside $SiN_x/SiO_2$ stack by means of RIE. The silicon substrate was then etched from the rear side by repeated deep RIE (Tegal 110 S/DE DRIE) steps to a depth of approximately 220 μm. The remaining silicon was removed by an anisotropic wet etch in 30% KOH at 70 °C, following a brief BHF dip to remove native oxide while preserving the front-side protective resist. Finally, the exposed $SiO_2$ was removed in BHF to release the free-standing membrane and open the nanopores. The completed devices were inspected by electrical *I–V* measurements and SEM.

**Electrical characterization and data processing**. Prior to electrical measurements, the nanopore chip was cleaned by oxygen plasma ashing at 1000 W for 5 min. It is then stored in a 1:1 vol. mixture of deionized water and ethanol. The chip was mounted in a custom-made polymethyl methacrylate flow cell and sealed on both sides using two polydimethylsiloxane O-rings with an inner diameter of 2 mm[29]. The KCl solutions were calibrated using a conductivity meter (Lab 945, Xylem Analytics Germany Sales GmbH & Co. KG). A pair of Ag/AgCl electrodes with a diameter of 0.75 mm (Warner Instruments LLC) was used to apply the bias voltage across the nanopore and record the ionic current. The electrical measurements were controlled using a patch-clamp amplifier (Axopatch 200B, Molecular Devices Inc.). The ionic current was digitized by an Axon Digidata 1550A system (Molecular Devices LLC) and recorded using Axon pCLAMP 12.2 software (Molecular Devices LLC). The entire experimental setup was placed inside a Faraday cage to minimize electromagnetic interference.

## Acknowledgment

This work was supported by the Swedish Research Council (Vetenskapsrådet) under Grant 2025-05333 to C.W. and 2022-03222 to S.-L. Z., Magnus Bergvalls Stiftelse in Sweden under Grant 2025-233 to C.W., Göran Gustafssons Stiftelse in Sweden with 1-year pris 2026 UU Fysik to C.W., and ÅForsk Foundation in Sweden under Grant 26-7 to C.W.

**Supporting Information**

# Memristive Behavior and Mechanism in Solid-State Nanopores

**Zhiwei Li[1], Ngan Hoang Pham[1,2*], Shi-Li Zhang[1], Chenyu Wen[1*]**

[1] *Division of Solid-State Electronics, Department of Electrical Engineering, Uppsala University, SE-751 21 Uppsala, Sweden*

[2] *Myfab Uppsala, Uppsala University, SE-751 21 Uppsala, Sweden*

*****Corresponding author:

ngan.pham@angstrom.uu.se

chenyu.wen@angstrom.uu.se

**Table of Contents:**

## Supporting Note 1: Decomposition of the measured I–V response

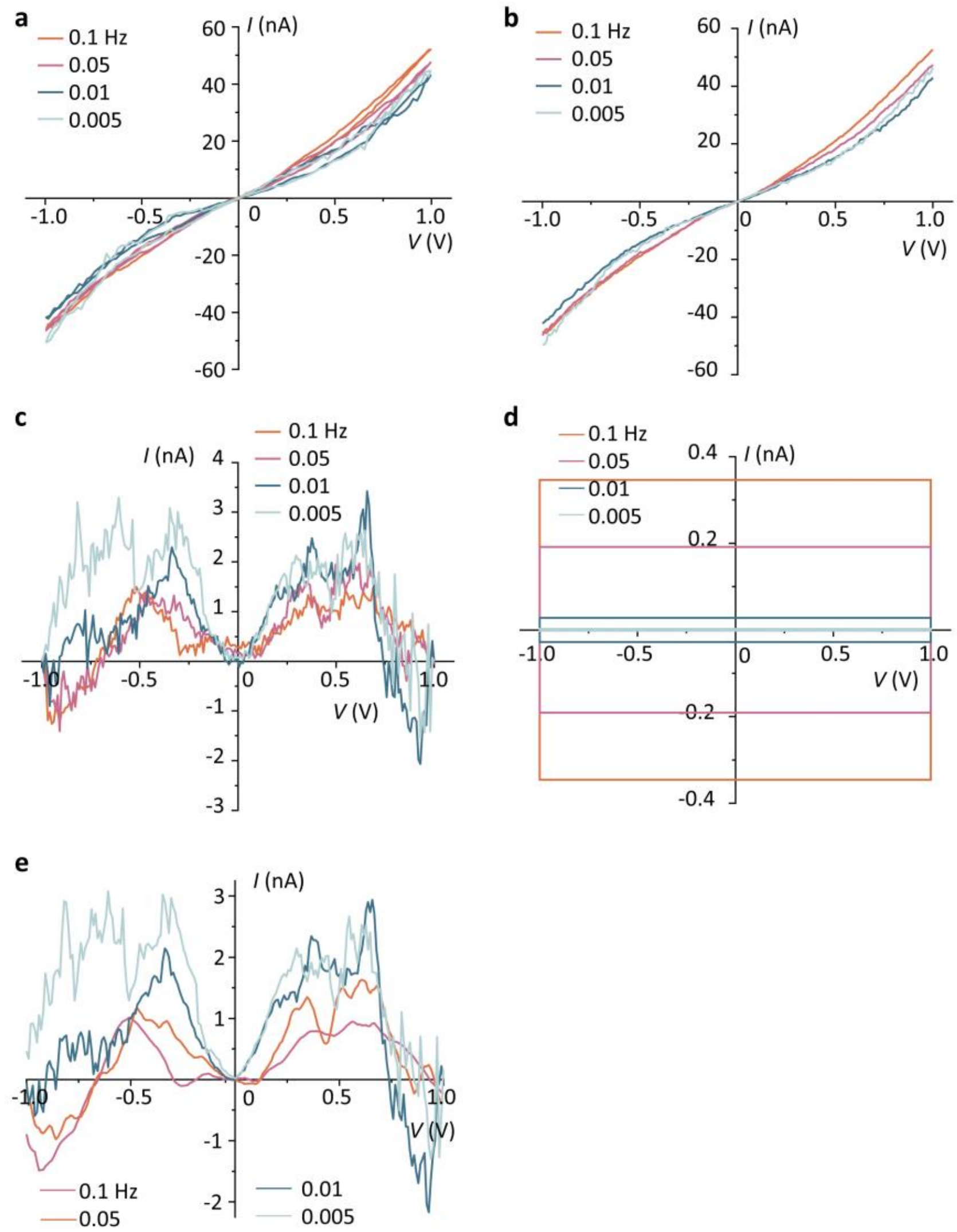

**Fig.S1.** Decomposition of the measured SSNP current. **a**, Original hysteretic *I*–*V* response under triangular voltage excitation. **b**, Extracted resistive component. **c**, Residual increasing-voltage branch after subtraction of the resistive component. **d**, Capacitive current determined from the residual current at zero applied voltage. **e**, Memristive current obtained after subtraction of both the resistive and capacitive components.

The measured solid-state nanopore (SSNP) response contains conductive, capacitive, and history-dependent contributions. Figure S1a shows the original hysteretic *I*–*V* curve obtained under triangular voltage excitation. To isolate the resistive component, the currents recorded during the increasing- and decreasing-voltage sweeps were evaluated at the same applied voltage and averaged. The resulting resistive component is shown in Fig. S1b.

Subtracting the extracted resistive current from the original increasing-voltage branch yields the residual current shown in Fig. S1c. This residual contains both capacitive and memristive components. Because the resistive current vanishes at zero applied voltage, the finite residual current at $V = 0$ was used to determine the capacitive current, as shown in Fig. S1d. Under triangular excitation, this contribution remains approximately constant during each linear voltage ramp and reverses sign when the sweeping direction changes.

The memristive current was finally obtained by further subtracting the reconstructed capacitive contribution. The resulting signal in Fig. S1e represents the memristive component used for subsequent kinetic modeling and hysteresis-area analysis.

**Supporting Note 2: Detailed description of the adsorption model**

The adsorption model is introduced as a minimal state-variable description of the isolated memristive current. It is not intended to identify a single elementary chemical reaction. Instead, it represents the net interfacial state that alters the surface charge density and hence the surface conductance of the solid-state nanopore. The fractional occupation of electrically active interfacial states is written as

$$\theta(t) = \frac{\Gamma(t)}{\Gamma_{max}}, \qquad 0 \leq \theta \leq 1 \qquad (S1)$$

where, $\Gamma(t)$ is the instantaneous surface density of occupied states and $\Gamma_{max}$ is their maximum effective density. The corresponding surface charge density is

$$\sigma(t) = \sigma_0 + zqN_A\Gamma_{\mathrm{max}}\theta(t) \qquad (S2)$$

where $\sigma_0$ is the intrinsic surface charge density obtained independently from the resistive-conductance model, $z$ the valence assigned to the effective adsorbed species, $q$ elementary charge, and $N_A$ Avogadro constant. The dynamic variable $\theta$ may represent reversible counterion association, changes in the protonation state of ionizable surface groups, or local reorganisation of the electrical double layer. These microscopic processes are not resolved individually and are instead represented by a common effective interfacial state.

This interpretation is consistent with the known chemistry of hydrated $SiN_x$ surfaces. Site-binding descriptions of $Si_3N_4$ interfaces include both silanol- and amine-related groups, and the resulting surface charge depends on $p$H, ionic strength, and the relative abundance of the two site types. Experiments on solid-state nanopores have also linked ionic-current fluctuations to time-dependent surface-charge regulation, while molecular-dynamics simulations have shown that delayed occupation of ion-binding sites can couple an internal ionic state to conductance and generate a memristive response[1–4].

The evolution of the effective surface coverage is described by Langmuir-type kinetics,

$$\frac{d\theta}{dt} = k_{\mathrm{ad}}(E)c[1-\theta] - k_{\mathrm{des}}(E)\theta \qquad (S3)$$

where $c$ is the electrolyte concentration, $k_{\mathrm{ad}}(E)$ is the field-dependent adsorption coefficient, and $k_{\mathrm{des}}(E)$ is the field-dependent desorption rate. When $c$ is in $mol\, m^{-3}$, $k_{\mathrm{ad}}$ has a unit of $\mathrm{m}^3\mathrm{mol}^{-1}s^{-1}$, and $k_{\mathrm{des}}$ has a unit of $\mathrm{s}^{-1}$.

The field-dependent rates are written as

$$k_{\mathrm{ad}}(E) = k_{\mathrm{ad,0}}e^{\beta_{\mathrm{a}}|E|+\beta_{\mathrm{s}}E} \qquad (S4)$$

$$k_{\mathrm{des}}(E) = k_{\mathrm{des,0}}e^{-\gamma_{\mathrm{a}}|E|-\gamma_{\mathrm{s}}E} \qquad (S5)$$

with

$$E = \frac{V}{L_{\mathrm{eff}}} \tag{S6}$$

Here, $L_{\mathrm{eff}}$ is the effective transport length rather than the nominal membrane thickness alone. It accounts approximately for the electric-field extension into the access regions. The local electric field around a nanopore is spatially non-uniform, and charged exterior membrane surfaces can contribute to ionic transport. Consequently, $E$ should be regarded as an effective axial electric field rather than the local electric field at an individual adsorption site[5,6].

The instantaneous surface conductance is

$$G_{\mathrm{s}}(\theta) = \frac{\pi D_p}{L_{\mathrm{eff}}} \mu |\sigma(t)| \tag{S7}$$

where $D_p$ is the pore diameter and $\mu$ is the mobility of the dominant counterion.

The parameter $\Gamma_{\max}$ determines the largest surface-charge modulation available to the effective interfacial state. It should not be interpreted as the total number of atoms or chemical groups on the physical surface. Only groups that can change their charge or occupancy within the experimental timescale contribute to the measured hysteresis. Inaccessible groups, permanently protonated or deprotonated sites, and sites whose kinetics are much faster or slower than the voltage period do not contribute appreciably to the extracted state variable[7].

As an order-of-magnitude benchmark, direct measurements on hydrated porous silica report hydroxyl densities of approximately 4.16–6.56 -OH groups per $nm^2$. $SiN_x$ surfaces are chemically distinct from pure silica, but their hydrated surface layer can contain silanol- and amine-related groups. A physically interpretable effective-state density of approximately 0.6–12 sites per $nm^2$, corresponding to

$$10^{-6}\ \mathrm{mol\ m^{-2}} \le \Gamma_{\max} \le 10^{-5}\ mol\ m^{-2}$$

is therefore appropriate for sensitivity analysis. This interval is deliberately broader than the hydroxyl density of pure silica because $\Gamma_{\max}$ represents a combined electrical state rather than a count of one chemically identified species[1,2,8].

For the reference geometry used in the current implementation,

$$D_{\mathrm{p}} = 15\ nm, L_{\mathrm{eff}} = 55nm$$

the fitted $\Gamma_{\max}$ is $1.37 \times 10^{-5}\ \mathrm{mol\ m^{-2}}$. This value lies within the physically motivated interval above.

The coefficient $k_{\mathrm{ad,0}}$ represents the zero-field probability per unit time and concentration that an available interfacial state becomes occupied. It includes transport towards the interface, partial dehydration, reorganisation of interfacial water, and the probability of crossing the effective adsorption barrier. The coefficient $k_{\mathrm{des,0}}$ represents the zero-field probability per unit time that an occupied state is released. These coefficients are effective kinetic parameters and should not be identified with elementary molecular collision frequencies.

Temporary ion binding and barrier-crossing descriptions have previously been used in kinetic models of nanopore transport. In such models, ions undergo Brownian motion and can be temporarily retained by traps or interfacial energy barriers. The use of positive exponential rate constants is consistent with barrier-crossing theory. Experimental measurements at charged interfaces further show that monovalent cation adsorption and desorption can occur over tens of seconds and can have substantially different rates because of changes in hydration and adsorbed-ion configuration[9,10].

At a fixed concentration, the experimental data identify the product $k_{\mathrm{ad},0}c$ not $k_{\mathrm{ad},0}$ alone. The optimiser therefore uses

$$ln(\ k_{\mathrm{ad},0}c)$$

as the fitted adsorption-rate parameter. Natural logarithms are used because that the rates are positive and can span several orders of magnitude.

At 1 M KCl, the reference fitted adsorption coefficient gives $\tau(0) \approx 43$ s, this relaxation time lies within the timescale sampled by the applied waveforms, whose periods range from 10 to 200 s. The model therefore operates in a regime where the surface state can evolve appreciably but cannot follow the voltage instantaneously. At 1 mM KCl, the same intrinsic adsorption coefficient gives $\tau(0) \approx 281$ s and a substantially smaller equilibrium coverage, illustrating how concentration alters both the accessible state and its relaxation.

The initial numerical fitting bounds were

$$-12 \leq \ln(\ k_{\mathrm{ad},0}c), ln(\ k_{des,0}) \leq 4$$

These are broad optimization bounds rather than expected physical intervals.

The parameters $\beta_{\mathrm{a}}$, $\beta_{\mathrm{s}}$, $\gamma_{\mathrm{a}}$, and $\gamma_{\mathrm{s}}$ describe how the effective adsorption and desorption barriers change with the applied field. The $|E|$ terms represent polarity-independent or common-mode changes. They may include field-enhanced interfacial polarization, ionic focusing, modification of the hydration environment, and changes in the probability of reaching an adsorption configuration. The signed $E$ terms describe polarity asymmetry arising from unequal pore entrances, non-uniform surface chemistry, asymmetric interfacial polarization, or differences between the two membrane surfaces.

The coefficients $\beta_{\mathrm{a}}$, $\beta_{\mathrm{s}}$, $\gamma_{\mathrm{a}}$, and $\gamma_{\mathrm{s}}$ have units of inverse electric field, with units of $(\mathrm{V\ m}^{-1})^{-1}$. Their signs determine the direction of the rate modulation. For the adsorption rate, a positive $\beta_{\mathrm{a}}$ increases adsorption with field magnitude, whereas a negative value suppresses it. A positive $\beta_{\mathrm{s}}$ favours adsorption under positive field and suppresses it under negative field. The interpretation of the $\gamma$ coefficients account for the minus signs in the desorption expression.

An exponential dependence is physically consistent with an activated process in which an applied field changes the electrostatic work along an effective reaction coordinate. Voltage-dependent ion-binding models similarly predict exponential changes in binding or dissociation when a charged species enters part of an electrical potential drop.

The initial numerical fitting bounds were

$$-0.549 \leq \beta_{\mathrm{a}}, \beta_{s}\ , \gamma_{a}, \gamma_{s} \leq 0.549\ (\mathrm{MV\ m}^{-1})^{-1}$$

The adsorption fit contains seven independent interfacial parameters, $\Gamma_{\max}$, $k_{\mathrm{ad},0}$, $k_{\mathrm{des},0}$, $\beta_{\mathrm{a}}$, $\beta_{\mathrm{s}}$, $\gamma_{\mathrm{a}}$, and $\gamma_{\mathrm{s}}$. The pore diameter $D_{\mathrm{p}}$, effective transport length $L_{\mathrm{eff}}$, ionic mobility $\mu$, electrolyte concentration $c$, valence $z$, and intrinsic surface charge density $\sigma_0$ are not fitting parameters in the adsorption model. The pore geometry is obtained from fabrication or conductance measurements, while $\sigma_0$ is obtained from the independently extracted resistive component. This separation reduces parameter correlation and prevents the adsorption model from compensating for uncertainties in the static pore conductance. The intrinsic surface charge density should also be treated as device- and condition-dependent. Experimental studies show that the charge density of $SiN_x$ nanopores depends on *p*H, salt concentration, pore geometry, and fabrication-dependent surface chemistry.

**Supporting Note 3: MNIST handwritten-digit recognition simulation**

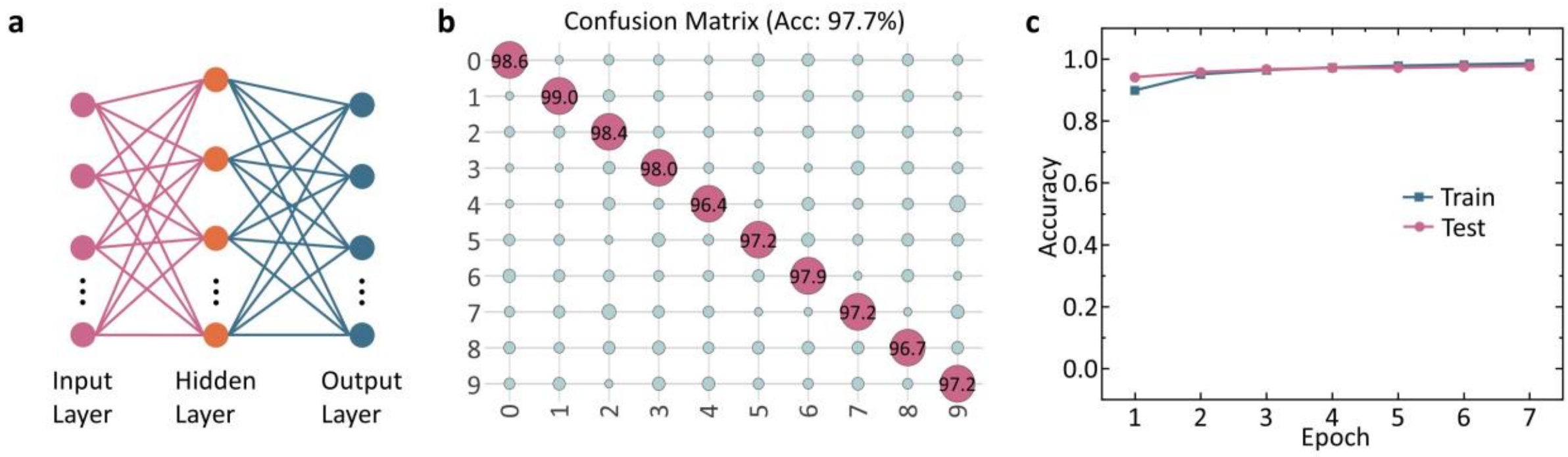

**Fig.S2.** MNIST handwritten-digit recognition simulation. **a**, Schematic of the fully connected neural network with a 784–256–10 architecture. **b**, Bubble map of the normalized confusion matrix for the 10,000-image MNIST test set. The overall test accuracy was 97.7%. **c**, Evolution of the training and test accuracies over seven training epochs.

To evaluate the potential of the measured memristive response for neuromorphic computing, a device-aware neural-network simulation was performed using the MNIST handwritten-digit dataset. The dataset contained 60,000 training images and 10,000 test images. Each 28×28-pixel image was normalized to the range of 0–1 and flattened into a 784-element input vector. A fully connected multilayer perceptron with a 784–256–10 architecture was employed, comprising 784 input neurons, 256 hidden neurons with rectified linear unit activation, and 10 output neurons corresponding to the digits 0–9.

The experimentally measured potentiation and depression responses each contained 15 states. The two branches were independently normalized to 0–1, the depression sequence was reversed, and the corresponding potentiation and depression states were averaged to obtain a common set of 15 nonuniform normalized conductance levels.

During quantization-aware training, the magnitude of every synaptic weight was projected onto the nearest experimentally derived state. The network was trained for seven epochs using the Adam optimizer. Cross-entropy loss was used for ten-class classification. The final device-quantized network achieved a test accuracy of 97.70%, compared with 97.68% when the corresponding floating-point weights were used. Comparable accuracies indicate that the experimentally derived 15-state response can support MNIST classification within the present algorithm-level model.

## Supporting Figure

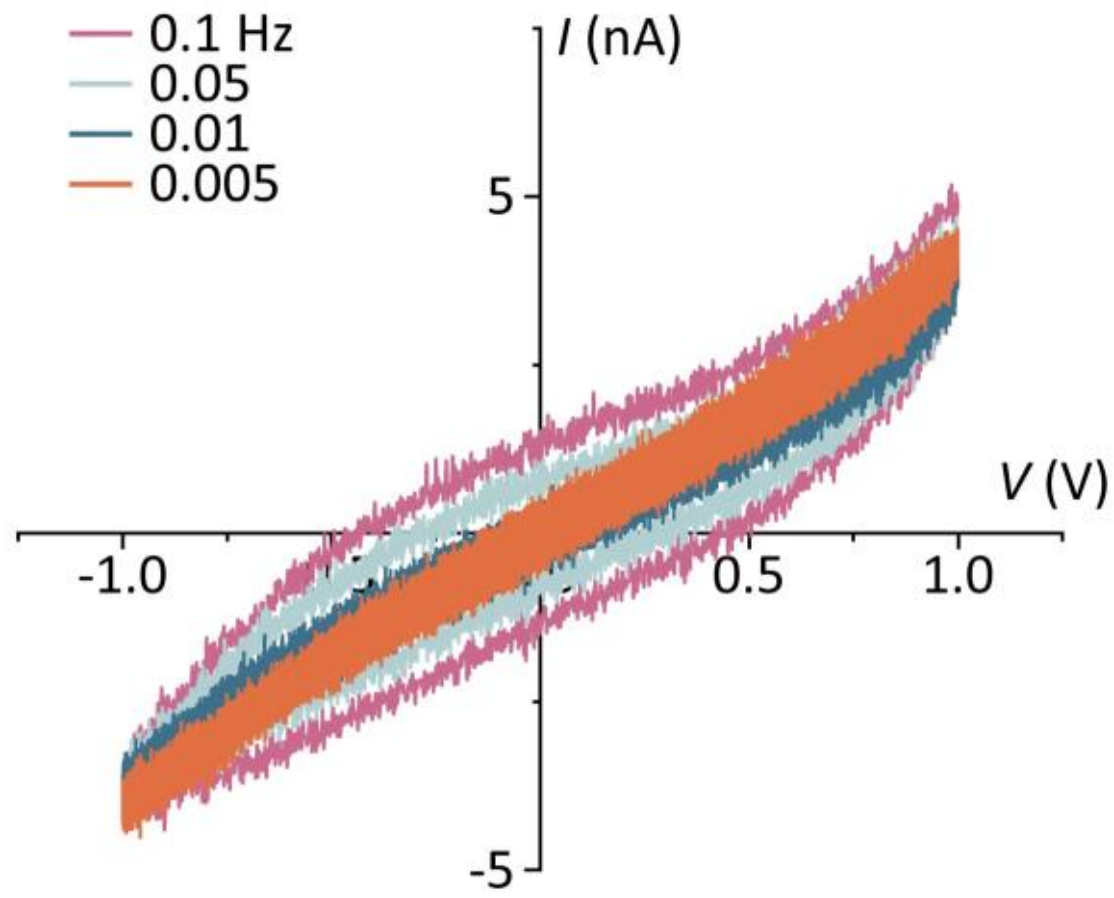

Fig. S3. *I–V* characteristics of a membrane-only control device without nanopore.

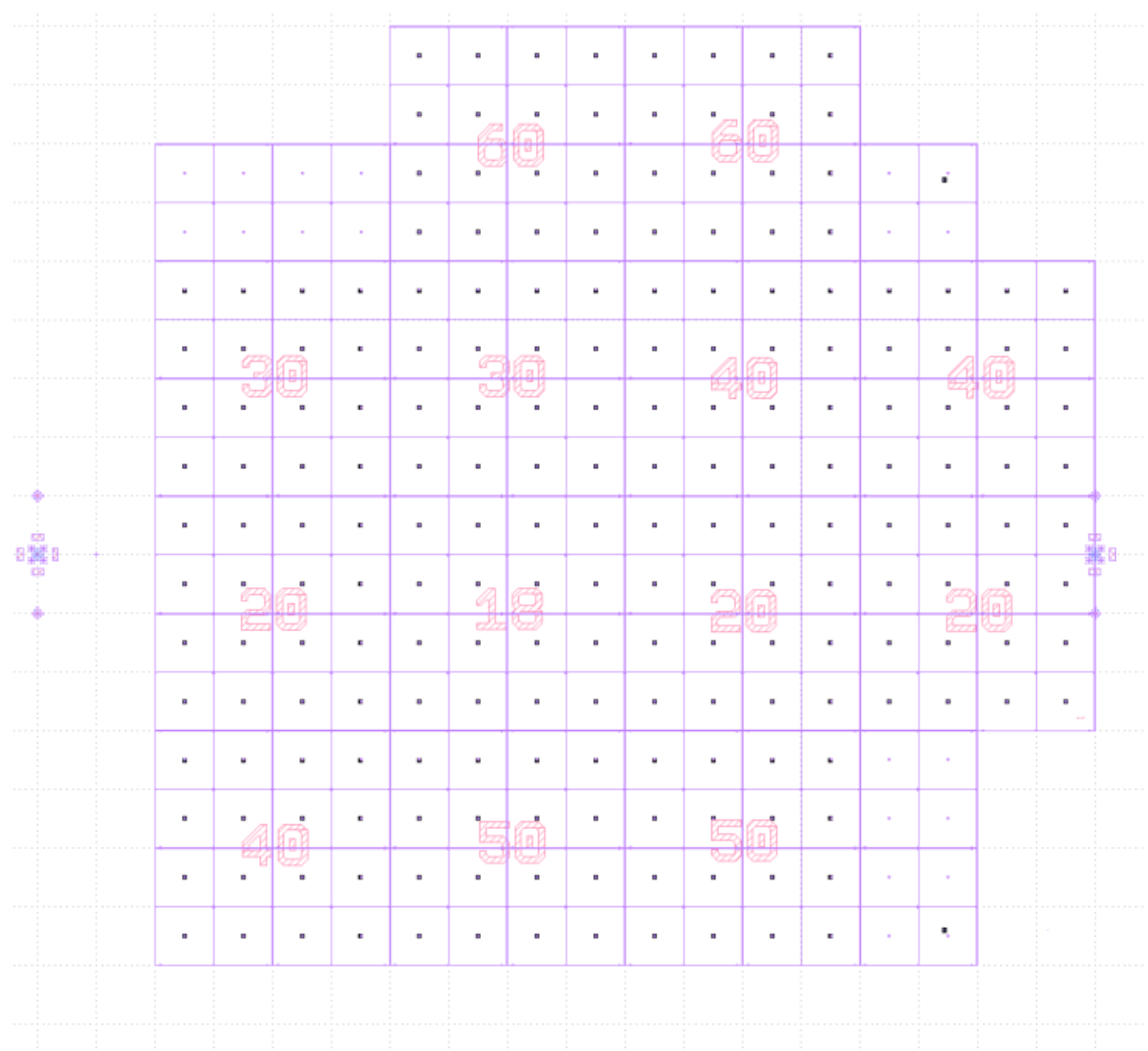

Fig. S4. Wafer-scale layout of the silicon nanopore devices. The yield is75%.

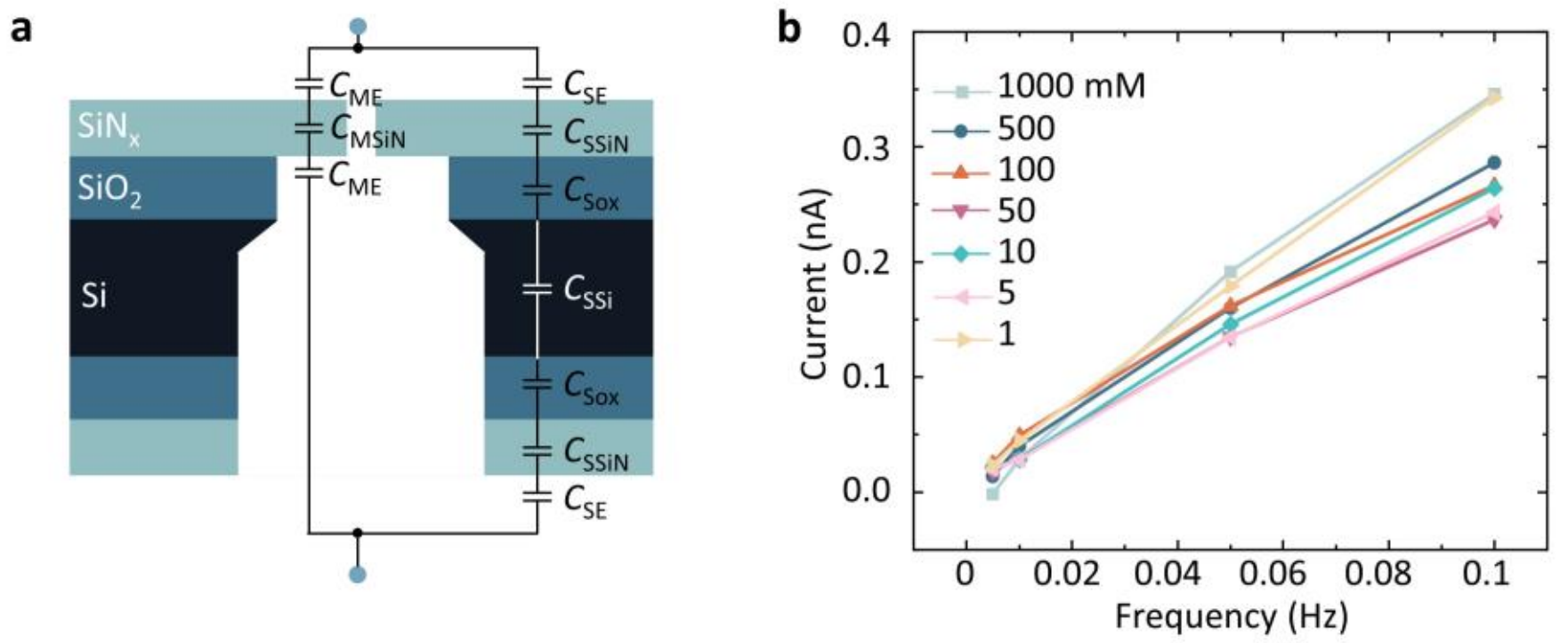

Fig. S5. Capacitive response of the SSNP device. **a**, Equivalent capacitance circuit of the free-standing membrane and substrate-supported regions, including the electrical double-layer, $SiN_x$, $SiO_2$, and silicon. **b,** Extracted capacitive current as a function of excitation frequency at different KCl concentrations.